\documentclass[runningheads]{llncs}

\usepackage[T1]{fontenc}
\usepackage{graphicx}
\usepackage{amsmath,amssymb,amsfonts}
\usepackage{booktabs}
\usepackage{microtype}
\usepackage{hyperref}

\usepackage{multirow}

\usepackage{tabularx}
\usepackage{cite}

\usepackage{xcolor}
\begin{document}

\title{NVV-Locator: From Transcript Tags to Acoustic Boundaries for Fine-Grained Nonverbal Vocalization Grounding}

\titlerunning{NVV-Locator}

\author{Yuang Cao\inst{1,4} \and
Bingshen Mu\inst{1} \and
Zhennan Lin\inst{1} \and
Guojian Li\inst{1} \and
Haoyue Zhan\inst{4} \and
Jie Liu\inst{4} \and
Chuan Xie\inst{4} \and
Qiang Zhang\inst{4} \and
Liumeng Xue\inst{2,3}\thanks{Corresponding authors.} \and
Lei Xie\inst{1}\textsuperscript{$\star$}}

\authorrunning{Y. Cao et al.}

% \institute{ASLP@NPU, Northwestern Polytechnical University, China
% \and
% Nanjing University, China
% \and
% Shanghai Lingguang Zhaxian Technology, China\\
% \email{yacao@mail.nwpu.edu.cn, lmxue@nju.edu.cn, lxie@nwpu.edu.cn}}

\institute{
ASLP@NPU, Northwestern Polytechnical University, China
\and
State Key Laboratory of Novel Software Technology, Nanjing University, China
\and
School of Intelligence Science and Technology, Nanjing University, China
\and
Shanghai Lingguang Zhaxian Technology, China\\
\email{yacao@mail.nwpu.edu.cn, lmxue@nju.edu.cn, lxie@nwpu.edu.cn}
}

\maketitle

% \vspace{-20pt}
\begin{abstract}
Human speech includes nonverbal vocalizations (NVVs), such as laughter, sighs, breaths, and coughs, which convey affective and interactional information. Existing approaches typically represent NVVs as transcript-level tags, providing limited supervision for their waveform-time boundaries. We present NVV-Locator for fine-grained NVV temporal grounding. We first unify 26 NVV categories across public resources and construct large-scale timestamp-supervised training data through dual-LLM verification, transcript-guided forced alignment, and energy-based boundary refinement. We further introduce NVV-TimeBench, an expert-refined benchmark with 667 utterances and 1,094 events. NVV-Locator uses a non-autoregressive slot-filling architecture to jointly predict lexical timestamps, NVV categories, and event boundaries. On NVV-TimeBench, it achieves 71.0\% Micro F1, 70.2\% Macro F1, 80.4\% Macro mIoU, and 59.6\,ms Macro mMAE, outperforming the evaluated large audio model counterparts. Evaluation on an external corpus further demonstrates the cross-corpus generalization of NVV-Locator.Our demo is available.\footnotemark

\footnotetext{Demo page: \url{https://nvv-locator.github.io/Demo-Page/}}
\end{abstract}


\section{Introduction}
Human spoken communication extends beyond lexical content. Nonverbal vocalizations (NVVs), such as laughter, sighs, breaths, and coughs, convey affective, attitudinal, and interactional information~\cite{borisov2025nonverbaltts,ye2025scalable,bai2026synparaspeech,mai2026mnv,DBLP:conf/mm/WuL0WLJBZL25,liao2025nvspeech,xue2026nvvsuperbench,yang2026beyond,ni2026nv}. Unlike lexical words, these events often lack stable linguistic forms but occupy specific acoustic intervals in continuous speech. Their accurate recognition and temporal localization are therefore important for speech understanding, expressive generation, and downstream audio editing.

Early NVV studies mainly formulate the problem as clip- or utterance-level acoustic event classification, as in ESC-50~\cite{DBLP:conf/mm/Piczak15} and VocalSound~\cite{DBLP:conf/icassp/GongYG22}, providing limited supervision for continuous-speech localization. More recent work has collected, synthesized, or annotated NVV corpora through automatic detect-then-align pipelines, including NonverbalTTS~\cite{borisov2025nonverbaltts} and  NonVerbalSpeech-38K~\cite{ye2025scalable}; synthesis- or editing-based approaches, such as SynParaSpeech~\cite{bai2026synparaspeech} and CapSpeech~\cite{wang2026capspeech}; and manually curated or hybrid resources, including MNV-17~\cite{mai2026mnv}, SMIIP-NV~\cite{DBLP:conf/mm/WuL0WLJBZL25}, and NVSpeech-170k~\cite{liao2025nvspeech}. These resources differ substantially in category inventories, scale, annotation precision, and acoustic naturalness.

Recent benchmarks have further advanced NVV generation and transcript-level understanding. NVV-SuperBench~\cite{xue2026nvvsuperbench} and NV-Bench~\cite{ni2026nv} evaluates controllable NVV generation, whereas WESR-Bench~\cite{yang2026wesr} evaluates NVV locations relative to the transcript. However, many existing approaches still represent NVVs primarily as discrete semantic tags, for example by inserting \texttt{[laughter]} into a transcript. Such representations capture event identity and textual context, but do not explicitly preserve acoustic onset, offset, or duration on the waveform timeline. Moreover, transcript-relative positions do not directly reflect the physical location of an NVV in the audio signal. Timestamped NVV data therefore remain scarce, while manual boundary annotation is expensive and difficult to scale.

Fine-grained waveform-time grounding also poses modeling challenges. Conventional forced-alignment systems, such as Montreal Forced Aligner (MFA)~\cite{DBLP:conf/interspeech/McAuliffeSM0S17} built on Kaldi~\cite{povey2011kaldi}, can estimate accurate lexical boundaries but rely on pronunciation lexicons and are difficult to extend to out-of-vocabulary NVVs. Neural systems, including NeMo~\cite{DBLP:conf/interspeech/RastorguevaLG23} and WhisperX~\cite{DBLP:conf/interspeech/BainHHZ23}, reduce lexicon dependence but are primarily designed for lexical speech and may yield degraded alignments when transcripts contain non-lexical events. Recent Speech-LLMs have also explored timestamp prediction; however, autoregressive timestamp generation does not inherently enforce temporal consistency. LLM-ForcedAligner~\cite{mu2026llm} addresses this issue through globally conditioned parallel boundary prediction, motivating a non-autoregressive formulation for joint lexical and NVV grounding.

To address these limitations, we formulate NVV temporal grounding as joint event recognition and waveform-time localization. We construct a unified 26-category taxonomy, develop a scalable automated pipeline for timestamp-supervised data construction, and curate NVV-TimeBench for fine-grained evaluation. We further propose NVV-Locator, a non-autoregressive slot-filling model that jointly predicts lexical timestamps, NVV categories, and NVV boundaries.

In summary, our main contributions are threefold:
\begin{itemize}
    \item We construct a unified paralinguistic taxonomy comprising 26 physical NVV categories, resolving annotation fragmentation across existing open corpora. We further develop a scalable automated pipeline for constructing a large-scale NVV dataset with timestamp supervision.

    \item We introduce NVV-TimeBench, a high-quality manually corrected benchmark for fine-grained NVV temporal grounding, with sub-20\,ms inter-annotator boundary discrepancy.

    \item We propose NVV-Locator, a lightweight non-autoregressive slot-filling model for joint lexical and NVV temporal grounding. By avoiding autoregressive timestamp decoding, it mitigates temporal inconsistency and achieves strong temporal grounding performance.

\end{itemize}

\section{Dataset Construction}

\subsection{A Unified Taxonomy of Nonverbal Vocalizations}
Existing NVV resources differ substantially in category inventories, naming conventions, and annotation granularity. Acoustically similar events may receive different labels across datasets, while functionally similar interjections may be split according to surface phonetic form. Such inconsistency hinders data aggregation, exacerbates sparsity for rare categories, and limits cross-corpus generalization.

We construct a unified taxonomy of 26 NVV categories by harmonizing four public resources: NVSpeech-170k~\cite{liao2025nvspeech}, MNV-17~\cite{mai2026mnv}, NVS-38K~\cite{ye2025scalable}, and SMIIP-NV~\cite{DBLP:conf/mm/WuL0WLJBZL25}. The taxonomy supports cross-corpus integration and fine-grained temporal grounding through two principles: acoustic normalization for physically defined vocal events and functional abstraction for lexicalized paralinguistic expressions.

\paragraph{Acoustic normalization for physical vocal events.}
For non-lexical events with identifiable acoustic or physiological realizations, synonymous or near-synonymous labels are merged into canonical categories. For example, \textit{[Throat Clearing]} combines \textit{[throatclearing]} from NVS-38K~\cite{ye2025scalable} and \textit{[ahem]} from MNV-17~\cite{mai2026mnv}, while \textit{[Sniff]} unifies \textit{[sniff]} and \textit{[sniffle]}. Similarly, \textit{[Laughter]}, \textit{[Cough]}, \textit{[Sigh]}, and \textit{[Breath]} are aligned across corpora when their source labels denote acoustically comparable events. This reduces unnecessary fragmentation while preserving distinctions between materially different acoustic phenomena.

\paragraph{Functional abstraction for lexicalized interjections.}
Some datasets distinguish interjections mainly by phonetic realization. NVSpeech-170k~\cite{liao2025nvspeech}, for example, separately annotates \textit{[Surprise-ah]}, \textit{[Surprise-oh]}, and \textit{[Surprise-wa]}, as well as \textit{[Question-en]} and \textit{[Question-ah]}. As these variants share pragmatic functions despite differing vowels or surface forms, we map them to \textit{[Surprise-X]} and \textit{[Question-X]}. The same strategy maps \textit{[Confirmation-en]} and \textit{[Dissatisfaction-hnn]} to \textit{[Confirmation-X]} and \textit{[Dissatisfaction-X]}, respectively. This reduces phonetic-variant sparsity and encourages functionally meaningful representations.

When available, original temporal boundaries are retained during data integration. We further add \textit{[Scream]}, \textit{[Roar]}, and \textit{[Burp]} to broaden taxonomy coverage beyond the source inventories. The resulting unified taxonomy is summarized in Table~\ref{tab:taxonomy}.

\begin{table}[!t]
\centering
\scriptsize
\caption{Unified 26-category NVV taxonomy and mappings from source datasets to canonical labels.}
\label{tab:taxonomy}
\begin{tabularx}{\textwidth}{
@{}
>{\raggedright\arraybackslash}p{0.16\textwidth}
@{\hspace{0.8em}}
>{\raggedright\arraybackslash}p{0.20\textwidth}
@{\hspace{0.8em}}
>{\raggedright\arraybackslash}X
@{}
}
\toprule
\textbf{Family} & \textbf{Canonical Category} & \textbf{Source Datasets and Original Labels} \\
\midrule
\multirow[t]{19}{0.16\textwidth}{\raggedright Physical Vocal Event}
& Sniff & \textbf{NVS-38K} ([sniff]); \textbf{MNV-17} ([sniffle]) \\
& Snore & \textbf{NVS-38K} ([snore]) \\
& Whistle & \textbf{MNV-17} ([whistle]) \\
& Hum & \textbf{MNV-17} ([hum]) \\
& Sneeze & \textbf{MNV-17} ([sneeze]) \\
& Yawn & \textbf{NVS-38K} ([yawn]) \\
& Moan & \textbf{MNV-17} ([moan]) \\
& Cry & \textbf{NVS-38K}, \textbf{SMIIP-NV} ([crying]); \textbf{NVSpeech-170k} ([Crying]) \\
& Cough & \textbf{NVS-38K} ([coughing]); \textbf{NVSpeech-170k} ([Cough]); \textbf{MNV-17}, \textbf{SMIIP-NV} ([cough]) \\
& Throat Clearing & \textbf{NVS-38K} ([throatclearing]); \textbf{MNV-17} ([ahem]) \\
& Sigh & \textbf{NVS-38K}, \textbf{MNV-17} ([sigh]); \textbf{NVSpeech-170k} ([Sigh]) \\
& Lip Smack & \textbf{MNV-17} ([smack]) \\
& Hiss & \textbf{MNV-17} ([hiss]) \\
& Gasp & \textbf{NVS-38K} ([gasp]) \\
& Laughter & \textbf{NVS-38K} ([laughing]); \textbf{NVSpeech-170k} ([Laughter]); \textbf{MNV-17} ([laugh], [chuckle]); \textbf{SMIIP-NV} ([laughter]) \\
& Breath & \textbf{NVS-38K} ([breath]); \textbf{NVSpeech-170k} ([Breathing]); \textbf{MNV-17} ([inhale/exhale/pant]) \\
& Scream & \textit{Added in this work} \\
& Roar & \textit{Added in this work} \\
& Burp & \textit{Added in this work} \\
\midrule
\multirow[t]{6}{0.16\textwidth}{\raggedright Functional Interjection}
& Shush & \textbf{NVSpeech-170k} ([Shh]) \\
& Confirmation-X & \textbf{NVSpeech-170k} ([Confirmation-en]) \\
& Dissatisfaction-X & \textbf{NVSpeech-170k} ([Dissatisfaction-hnn]) \\
& Question-X & \textbf{NVSpeech-170k} ([Question-en/ah/oh/ei/yi]) \\
& Filled Pause & \textbf{NVSpeech-170k} ([Uhm]) \\
& Surprise-X & \textbf{NVSpeech-170k} ([Surprise-ah/oh/wa/yo]) \\
\midrule
Associated Non-Speech Event
& Applause & \textbf{MNV-17} ([clap], [applaud]) \\
\bottomrule
\end{tabularx}
\end{table}

\subsection{Automatic Data Construction Pipeline}
\begin{figure}[htbp]
\centering
\includegraphics[width=\textwidth]{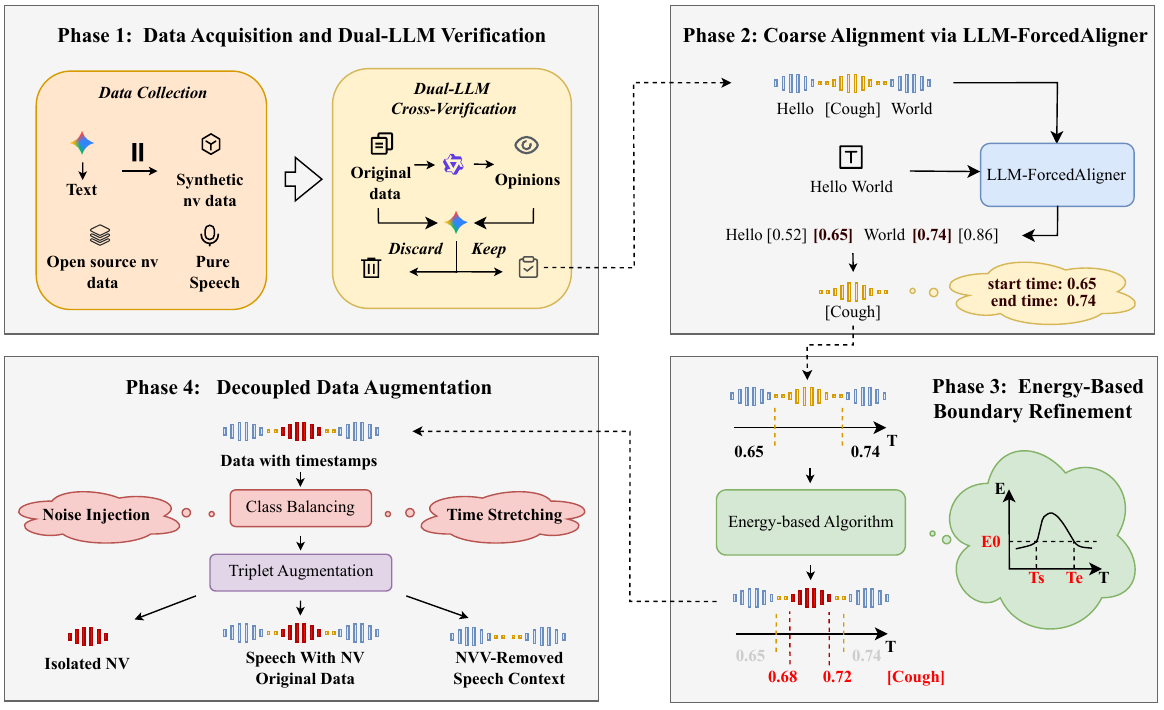}
\vspace{-5pt}
\caption{The proposed four-phase automated data production pipeline.
}
\vspace{10pt}
\label{fig:pipeline}
\end{figure}

To construct large-scale training data for transcript-conditioned NVV temporal grounding under the unified taxonomy, we develop a four-stage automatic data construction pipeline. Figure~\ref{fig:pipeline} illustrates an overview of the pipeline.

\subsubsection*{Phase 1: Candidate Data Acquisition and Dual-LLM Verification}

We acquire candidate training instances from three complementary sources: open-source resources containing NVV annotations, targeted synthetic NVV data for categories with limited coverage, and speech-only recordings used as negative controls. Because source annotations may contain background interference, overlapping speech, category ambiguity, or transcript--audio mismatch, candidate NVV instances are subjected to a sequential dual-LLM verification procedure.

Specifically, Qwen3-Omni~\cite{xu2025qwen3} first analyzes each audio--label pair and produces an initial diagnostic report concerning NVV presence, category consistency, and potential ambiguity. Gemini-2.5 subsequently reviews the original audio and candidate annotation, with the first-stage report provided as auxiliary evidence, and makes the final binary keep/discard decision. An instance is retained only when the target NVV is acoustically present, consistent with its assigned category, and sufficiently unambiguous for temporal grounding. Human spot checks are further conducted on a randomly sampled subset to audit the quality of the automated verification process.

\subsubsection*{Phase 2: Transcript-Guided Coarse Alignment via LLM-ForcedAligner}

For each retained utterance, we remove NVV tags from the corresponding transcription to obtain a lexical transcript. We then apply LLM-ForcedAligner~\cite{mu2026llm}  to align the lexical transcript with the audio and estimate character-level timestamps for Chinese and word-level timestamps for English.

For an NVV tag located between two adjacent lexical units, we construct a coarse bracketing window using the end time of the preceding unit and the start time of the succeeding unit $\mathcal{W} = [t_{\mathrm{left}}^{e}, t_{\mathrm{right}}^{s}]$, where $t_{\mathrm{left}}^{e}$ denotes the end time of the preceding lexical unit and $t_{\mathrm{right}}^{s}$ denotes the start time of the succeeding lexical unit. This transcript-guided interval provides a candidate region for the target NVV event; it is not treated as its final acoustic boundary.

\subsubsection*{Phase 3: Energy-Based Boundary Refinement}

The coarse bracketing window may include leading or trailing silence, residual background noise, and non-NVV acoustic content. We therefore refine the boundary estimates using the local energy profile within $\mathcal{W}$. Specifically, we compute an RMS-based short-time energy contour using a 5\,ms analysis window and smooth it with a 10\,ms moving window.

Let $E(t)$ denote the smoothed RMS energy at time $t$. We define an adaptive energy threshold as $\tau = \alpha \max_{t \in \mathcal{W}} E(t)$, where $\alpha = 0.15$. Starting from the two edges of $\mathcal{W}$, the algorithm removes low-energy regions until the smoothed energy reaches the threshold $\tau$. The resulting interval $[\hat{t}{s}, \hat{t}{e}]$ serves as the acoustically refined onset--offset estimate for the target NVV event. This procedure reduces the mismatch between transcript-derived lexical gaps and the temporal extent of the corresponding acoustic event.

\subsubsection*{Phase 4: Class Balancing and Decoupled Data Augmentation}

To mitigate class imbalance and improve discrimination between lexical speech and NVVs, we perform class balancing and decoupled data augmentation. For underrepresented categories, we apply noise injection and time stretching to increase acoustic diversity. When time stretching is applied, the associated onset and offset annotations are transformed using the same temporal scaling factor.

Each refined utterance is subsequently represented in three complementary views: (i) the original mixed speech recording containing the NVV event, (ii) an isolated NVV segment cropped using the refined timestamps, and (iii) an NVV-removed speech context obtained by removing the localized NVV interval from the original recording. These views provide mixed-context examples, event-focused examples, and negative-control speech contexts, respectively. The resulting data encourage the downstream model to distinguish acoustically grounded NVV events from lexical speech and speech-adjacent acoustic variation.

\subsection{NVV-TimeBench: An Expert-Refined Benchmark for Fine-Grained NVV Temporal Grounding}
We introduce NVV-TimeBench, an expert-refined benchmark for fine-grained NVV temporal grounding. Each instance contains an audio recording, its lexical transcript, and one or more annotations $(c,t_s,t_e)$, where $c$ denotes the canonical NVV category and $t_s$ and $t_e$ denote its acoustic onset and offset. The benchmark evaluates both event-category recognition and waveform-time boundary localization, rather than transcript-relative tag placement alone.

The benchmark is curated from public resources and generated data that are excluded from the model training set. Expert annotation and review are conducted throughout the curation process; a randomly selected subset is independently annotated by five experts, with two additional validators performing quality checks.

The final benchmark contains 667 utterances with 1,094 target events, totaling approximately 112.6 minutes of audio and covering all 26 categories in the unified taxonomy. Generic background-noise labels are excluded because they do not define a target event under the present grounding formulation. Category counts range from 22 to 159 instances, covering rare events such as \texttt{[Scream]}, \texttt{[Roar]}, and \texttt{[Burp]}, as well as frequent events including \texttt{[Laughter]} and \texttt{[Breath]}; NVV-TimeBench is therefore category-diverse rather than strictly class-balanced. Event durations have a mean of 0.52~s and a median of 0.40~s, spanning brief transients and sustained vocalizations of up to nearly 8~s.

\section{NVV-Locator: A Transcript-Conditioned Model for Fine-Grained NVV Temporal Grounding}

We propose \textbf{NVV-Locator}, a transcript-conditioned model for fine-grained temporal grounding of nonverbal vocalizations (NVVs) and a dedicated baseline for NVV-TimeBench. Given a speech recording and its lexical transcript, with all NVV annotations removed, NVV-Locator jointly estimates the temporal boundaries of lexical units and identifies locally associated NVV events, including their categories and acoustic onset and offset times. The lexical transcript is treated as an observed input rather than a generation target, allowing the model to focus its capacity on multimodal temporal alignment.

\begin{figure}[htbp]
\centering
\includegraphics[width=\textwidth]{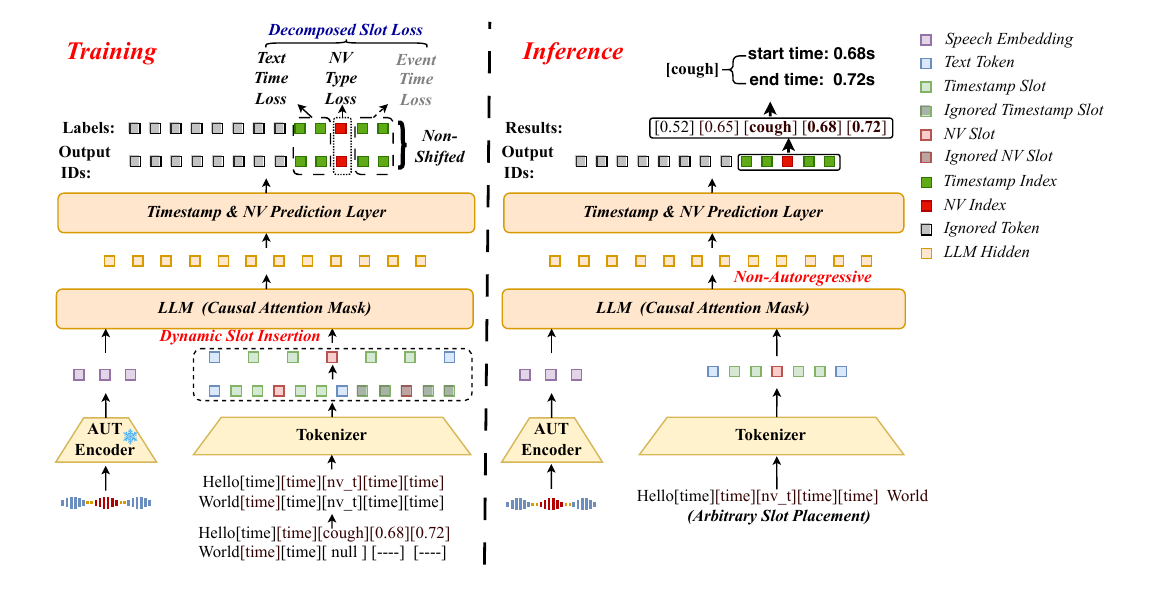}
\vspace{-15pt}
\caption{Architecture of NVV-Locator. A five-slot template is inserted at transcript positions to encode lexical start and end times, NVV type, and NVV onset and offset times. During training, the slot labels are supervised directly at their corresponding positions without next-token shifting. During inference, all slot labels are predicted in parallel and decoded into lexical boundaries and NVV temporal intervals.}
\label{fig:model_arch}
\end{figure}

\subsection{Non-Autoregressive Slot-Filling Architecture}
NVV-Locator is built on Qwen3-ASR-0.6B~\cite{shi2026qwen3}, comprising a pretrained audio encoder and a 0.6B-parameter language-model decoder. It reformulates continuous temporal grounding as structured slot filling over a transcript-aligned sequence.

Autoregressive timestamp generation can propagate errors across the output sequence and increases inference cost with the number of predicted boundaries. NVV-Locator instead predicts predefined temporal and event slots in parallel. Although the decoder retains the causal attention mask inherited from Qwen3-ASR, labels for all inserted slots are predicted simultaneously rather than generated recursively.

Let $x$ denote an input waveform and $\mathbf{H}_A=\mathrm{Enc}_A(x)$ the acoustic feature sequence. Given a lexical transcript
$
\mathbf{W}=(w_1,w_2,\ldots,w_N),
$,
where each $w_i$ is a Chinese character or English word, we insert a five-slot template after each transcript unit:
$
\mathcal{S}_i=
\langle
c_s^{(i)},c_e^{(i)},nv_t^{(i)},nv_s^{(i)},nv_e^{(i)}
\rangle$.
Here, $c_s^{(i)}$ and $c_e^{(i)}$ denote the start and end times of $w_i$, while $nv_t^{(i)}$, $nv_s^{(i)}$, and $nv_e^{(i)}$ denote the category, onset, and offset of the NVV associated with this position. When no NVV is present, $nv_t^{(i)}$ is assigned \texttt{[Null]}, and its timestamp slots are masked from supervision.

The resulting slot-augmented transcript is denoted by $\widetilde{\mathbf{W}}$. Dense lexical timestamp slots and sparse NVV slots jointly establish a shared temporal reference frame. Continuous time is discretized into $B=750$ bins with resolution $\Delta t=40$ ms, covering audio up to 30 seconds. The tokenizer is augmented with 778 task-specific output tokens: 750 timestamp-index tokens, 26 NVV-category tokens, one \texttt{[Null]} token, and one reserved background token not used for supervision.

Given $\mathbf{H}_A$ and $\widetilde{\mathbf{W}}$, the decoder produces representations for all inserted slots in one forward pass:
\begin{equation}
\mathbf{Z}=
\mathrm{LLM}_{\theta}
\left(
\mathbf{H}_A,\widetilde{\mathbf{W}}
\right).
\end{equation}
Let $\mathcal{R}=\{c_s,c_e,nv_t,nv_s,nv_e\}$ denote the slot types. The output distribution is factorized over slot positions:
\begin{equation}
P_{\theta}
\left(
\mathbf{Y}
\mid
\mathbf{H}_A,\widetilde{\mathbf{W}}
\right)
=
\prod_{i=1}^{N}
\prod_{r\in\mathcal{R}}
P_{\theta}
\left(
y_{i,r}
\mid
\mathbf{z}_{i,r}
\right),
\end{equation}
where $\mathbf{z}_{i,r}$ is the representation of slot $r$ at position $i$. Timestamp slots predict one of the $B$ time bins, whereas NVV-type slots predict an NVV category or \texttt{[Null]}.

Unlike standard next-token language modeling, supervision is applied directly at each slot without label shifting. All lexical and NVV boundaries are thus inferred in parallel, and soft expectation decoding over time-bin posteriors yields continuous timestamps with sub-bin precision.

\subsection{Noise-Robust Training and Optimization Strategies}
Because the training corpus is constructed through an automated large-scale pipeline, the resulting supervision may contain residual boundary noise. In addition, NVV events are sparse relative to lexical units, resulting in substantial imbalance between dense lexical timestamp labels and sparse event-level labels. To improve robustness under these conditions, we freeze the audio encoder during training and fine-tune the language-model decoder together with the timestamp and NVV prediction layer. This constrains adaptation to the multimodal alignment and slot-prediction components while reducing overfitting to noisy acoustic supervision.

\subsubsection*{Decomposed Slot Optimization and Conditional Masking}
We optimize the Timestamp \& NVV Prediction Layer using three complementary objectives: lexical timestamp loss $\mathcal{L}_{\text{text\_time}}$, NVV type loss $\mathcal{L}_{\text{nv\_type}}$, and NVV timestamp loss $\mathcal{L}_{\text{event\_time}}$:

\begin{equation}
\mathcal{L}_{\text{total}} = \mathcal{L}_{\text{text\_time}} + \mathcal{L}_{\text{nv\_type}} + \mathcal{L}_{\text{event\_time}}
\end{equation}

Each term is computed as the mean cross-entropy over its valid supervision positions. Specifically, $\mathcal{L}_{\text{text\_time}}$ is evaluated on valid lexical timestamp slots, while $\mathcal{L}_{\text{nv\_type}}$ is evaluated on all valid NVV-type slots, including the \texttt{[Null]} class. This allows the model to learn both NVV category recognition and the absence of an NVV event.

For NVV timestamps, supervision is meaningful only when an NVV is present. Let
\begin{equation}
\Omega_{\text{NV}}
=
\left\{
i
\mid
y^{(i)}_{nv_t}
\neq
\texttt{[Null]}
\right\}
\end{equation}
denote the set of transcript positions associated with valid NVV events. The event timestamp loss is defined as
\begin{equation}
\mathcal{L}_{\text{event\_time}}
=
\frac{1}{2|\Omega_{\text{NV}}|}
\sum_{i\in\Omega_{\text{NV}}}
\left[
\mathrm{CE}
\left(
\hat{y}^{(i)}_{nv_s},
y^{(i)}_{nv_s}
\right)
+
\mathrm{CE}
\left(
\hat{y}^{(i)}_{nv_e},
y^{(i)}_{nv_e}
\right)
\right].
\end{equation}

Here, the subscripts $s$ and $e$ denote the onset (start) and offset (end)
of an NVV event, respectively. $y^{(i)}_{nv_s}$ and $y^{(i)}_{nv_e}$ are
the ground-truth discrete time-bin labels for the onset and offset of the
NVV associated with position $i$. Correspondingly,
$\hat{y}^{(i)}_{nv_s}$ and $\hat{y}^{(i)}_{nv_e}$ denote the model's
predicted categorical distributions over the $B$ timestamp bins. Thus, the
two cross-entropy terms supervise onset and offset prediction separately.
The factor of $2$ averages the loss over the two temporal boundaries of each
valid NVV event.

For mini-batches without valid NVV events, $\mathcal{L}_{\text{event\_time}}$ is set to zero. This conditional masking prevents the model from learning arbitrary timestamps for non-existent events and ensures that event-boundary gradients originate only from acoustically grounded NVV instances. The per-term normalization also prevents dense lexical timestamp supervision from numerically dominating the sparse NVV-specific objectives.

\subsubsection*{Event-Level Hungarian Matching Validation}

Token-level cross-entropy does not directly assess utterance-level event correspondence. We therefore reserve the validation set and use category-aware Hungarian matching for checkpoint selection.

Let $\mathcal{E}^* = \{e^*_1, ..., e^*_U\}$ and $\hat{\mathcal{E}} = \{\hat{e}_1, ..., \hat{e}_V\}$ denote the ground-truth and predicted NVV events, respectively, each with a category, onset, and offset. Their temporal Intersection over Union (t-IoU) is
\begin{equation}
\text{t-IoU}(e^*,\hat{e})
=
\frac{\ell_{\cap}(e^*,\hat{e})}
{\ell_{\cup}(e^*,\hat{e})},
\end{equation}
where
$
\ell_{\cap}(e^*,\hat{e})
=
\max
\left(
0,
\min(t_e^*,\hat{t}_e)
-
\max(t_s^*,\hat{t}_s)
\right),
$
and
$
\ell_{\cup}(e^*,\hat{e})
=
(t_e^*-t_s^*)
+
(\hat{t}_e-\hat{t}_s)
-
\ell_{\cap}(e^*,\hat{e}).
$
We define the category-aware matching cost as
\begin{equation}
C_{i,j}
=
\begin{cases}
1-\text{t-IoU}(e_i^*,\hat{e}_j),
&
\text{if }
c_i^*=\hat{c}_j
\ \text{and}\
\text{t-IoU}(e_i^*,\hat{e}_j)\geq\tau_{\text{val}},
\\
+\infty,
&
\text{otherwise},
\end{cases}
\end{equation}
where $c_i^*$ and $\hat{c}_j$ are the ground-truth and predicted categories. We set $\tau_{\text{val}}=0.3$, a more permissive threshold than the $\text{t-IoU}\geq0.5$ criterion used for final benchmarking.

We augment the bipartite graph with dummy nodes and obtain the minimum-cost one-to-one assignment using the Hungarian algorithm. Matched pairs are counted as true positives, while unmatched predictions and references are counted as false positives and false negatives. We monitor Macro $F_1$ and boundary Mean Absolute Error (MAE) over matched pairs:
\begin{equation}
\mathrm{MAE}
=
\frac{1}{2|\mathcal{M}|}
\sum_{(i,j)\in\mathcal{M}}
\left(
|t_{s,i}^{*}-\hat{t}_{s,j}|
+
|t_{e,i}^{*}-\hat{t}_{e,j}|
\right),
\end{equation}
where $\mathcal{M}$ denotes the matched real-event pairs. This protocol selects checkpoints according to event-level temporal grounding quality rather than token-level accuracy alone.

\section{Experiments and Results}

\subsection{Experimental Setup}

\subsubsection*{Dataset}
Following the first three phases of the automated data construction pipeline, a total of 283.20 hours of candidate audio data are yielded. From this initial corpus, approximately 7.49 hours are partitioned for validation and evaluation, while the remaining 275.71 hours are subjected to Phase 4 (decoupled data augmentation), effectively expanding the final training set to 551.42 hours. Within the 7.49-hour validation split, 1.88 hours are allocated for expert manual refinement to construct the high-precision evaluation benchmark, whereas 5.61 hours serve as the development validation set for model selection and checkpoint optimization during training.

\subsubsection*{Baselines}
We evaluate NVV-Locator on NVV-TimeBench, the expert-refined benchmark for fine-grained NVV temporal grounding. We compare it with four representative large audio models (LAMs): Gemini-2.5-Pro~\cite{comanici2025gemini}, Qwen3-Omni-Instruct (30B)~\cite{xu2025qwen3}, Step-Audio-R1.1~\cite{tian2025step}, and MOSS-Audio-8B-Instruct~\cite{yang2026moss}. The LAMs are evaluated zero-shot and prompted to output canonical NVV categories together with their onset and offset times. Their outputs are mapped to the unified 26-category taxonomy and evaluated using the same protocol as NVV-Locator.

\subsubsection*{Evaluation Metrics}
To evaluate NVV event recognition and temporal boundary accuracy, we adopt three complementary metrics. Event-Level F1 (t-IoU $\geq 0.5$) is the primary measure of event identification under a temporal matching criterion. A prediction is counted as a true positive only when its predicted NVV category matches the ground-truth category and the temporal Intersection over Union (t-IoU) between the predicted and reference intervals is at least 0.5. Mean absolute error (mMAE) measures boundary precision by averaging the absolute onset and offset errors over successfully matched pairs, reported in milliseconds. Mean IoU (mIoU) is computed over the same matched pairs as an auxiliary measure of temporal overlap.

Both micro- and macro-averaged results are reported. The \textbf{Micro average} pools true positives, false positives, and false negatives across all categories to derive global Precision, Recall, and F1; micro mIoU and mMAE are averaged over all matched pairs. The \textbf{Macro average} for F1 is the arithmetic mean of category-wise F1 scores. For mIoU and mMAE, macro averages are computed only over categories with at least one successfully matched pair (t-IoU $\geq 0.5$), thereby avoiding undefined boundary statistics for categories without matched events.

\subsection{Main Results}
Table~\ref{tab:main_results} summarizes the results on NVV-TimeBench. Among the LAM baselines, Gemini-2.5-Pro achieves the strongest aggregate performance, with a Macro F1 of 49.8\%, a Macro mIoU of 72.8\%, and a Macro mMAE of 92.6\,ms. Qwen3-Omni, Step-Audio, and MOSS-Audio obtain substantially lower Macro F1 scores of 32.9\%, 27.9\%, and 12.5\%, respectively. These results indicate that, despite occasional accurate matches for individual event types, current general-purpose LAMs remain limited in reliably identifying and temporally grounding fine-grained NVV events.

NVV-Locator achieves the best aggregate performance across all three metrics. It attains a Micro F1 of 71.0\% and a Macro F1 of 70.2\%, improving over the strongest baseline by 25.2 and 20.4 absolute percentage points, respectively. It further achieves a Macro mIoU of 80.4\% and a Macro mMAE of 59.6\,ms, corresponding to a 7.6-point gain in temporal overlap and a 33.0\,ms reduction in boundary error relative to Gemini-2.5-Pro. Together, these results demonstrate the effectiveness of NVV-Locator for fine-grained NVV event identification and temporal boundary localization.

\begin{table}[tbp]
\centering
\caption{Main results on NVV-TimeBench. Each cell reports Event-Level F1 $\uparrow$ / mIoU $\uparrow$ / mMAE $\downarrow$, where F1 (\%) is computed with t-IoU $\geq 0.5$, and mIoU (\%) and mMAE (ms) are computed over successfully matched event pairs. Micro and Macro averages are given in the last two rows. \textbf{Bold} and \underline{underlining} denote the best and second-best values, respectively, for each metric within a row.}
\vspace{-3pt}
\label{tab:main_results}
\setlength{\tabcolsep}{8pt}
\newcommand{\three}[3]{\makebox[2.5em][r]{#1}\,/\,\makebox[2.5em][r]{#2}\,/\,\makebox[2.5em][r]{#3}}
\resizebox{\linewidth}{!}{%
\begin{tabular}{@{} l c c c c c @{}}
\toprule
\multirow{2}{*}{\textbf{Tag}} & \textbf{Gemini-2.5-Pro} & \textbf{Qwen3-Omni} & \textbf{Step-Audio} & \textbf{MOSS-Audio} & \textbf{NVV-Locator (Ours)} \\
& {\footnotesize F1 / mIoU / mMAE}
& {\footnotesize F1 / mIoU / mMAE}
& {\footnotesize F1 / mIoU / mMAE}
& {\footnotesize F1 / mIoU / mMAE}
& {\footnotesize F1 / mIoU / mMAE} \\
\midrule
\texttt{Applause}         & \three{\underline{32.7}}{74.9}{91.0} & \three{17.6}{61.9}{222.0} & \three{13.3}{64.3}{154.3} & \three{5.7}{\textbf{91.8}}{\textbf{44.0}}  & \three{\textbf{90.6}}{\underline{78.0}}{\underline{81.1}} \\
\texttt{Breath}           & \three{18.8}{69.1}{84.4}  & \three{4.9}{59.4}{260.2}  & \three{\underline{30.1}}{66.9}{\underline{75.4}}  & \three{17.5}{\underline{69.3}}{\textbf{62.0}}  & \three{\textbf{47.7}}{\textbf{77.7}}{\textbf{62.0}} \\
\texttt{Burp}             & \three{\textbf{62.1}}{\underline{73.1}}{\underline{48.6}}  & \three{30.6}{63.7}{79.4}  & \three{26.2}{64.0}{58.9}  & \three{0.0}{--}{--}     & \three{\underline{59.1}}{\textbf{83.6}}{\textbf{30.4}} \\
\texttt{Confirmation-X}   & \three{\underline{27.7}}{74.7}{56.0}  & \three{8.1}{\textbf{81.4}}{\textbf{23.7}}   & \three{22.2}{70.2}{62.5}  & \three{2.9}{74.0}{65.0}   & \three{\textbf{48.5}}{\underline{79.0}}{\underline{40.5}} \\
\texttt{Cough}            & \three{\underline{41.9}}{66.6}{105.9} & \three{36.8}{71.5}{140.6} & \three{32.3}{\underline{75.0}}{\underline{93.3}}  & \three{35.8}{72.4}{95.9}  & \three{\textbf{78.1}}{\textbf{83.0}}{\textbf{50.0}} \\
\texttt{Cry}              & \three{64.9}{75.2}{240.9} & \three{\underline{70.3}}{\underline{77.5}}{\underline{200.9}} & \three{56.5}{75.6}{285.3} & \three{30.7}{66.1}{296.0} & \three{\textbf{75.5}}{\textbf{83.5}}{\textbf{129.9}} \\
\texttt{Dissatisfaction-X}& \three{18.9}{\underline{71.7}}{85.1}  & \three{\underline{27.5}}{71.5}{\underline{51.1}}  & \three{12.3}{65.9}{60.9}  & \three{0.0}{--}{--}     & \three{\textbf{82.8}}{\textbf{73.7}}{\textbf{39.9}} \\
\texttt{Filled Pause}     & \three{\underline{50.9}}{\underline{76.5}}{\underline{66.9}}  & \three{17.6}{65.7}{74.0}  & \three{26.8}{68.9}{84.5}  & \three{0.0}{--}{--}     & \three{\textbf{76.5}}{\textbf{87.9}}{\textbf{29.1}} \\
\texttt{Gasp}             & \three{\textbf{32.5}}{64.7}{73.8}  & \three{\underline{26.7}}{\underline{70.7}}{\underline{64.4}}  & \three{12.9}{\textbf{72.3}}{\textbf{63.1}}  & \three{20.4}{65.5}{82.3}  & \three{16.7}{62.3}{98.0} \\
\texttt{Hiss}             & \three{\underline{49.2}}{\underline{74.7}}{58.5}  & \three{9.3}{71.1}{\underline{55.7}}   & \three{0.0}{--}{--}     & \three{0.0}{--}{--}     & \three{\textbf{58.6}}{\textbf{77.0}}{\textbf{51.9}} \\
\texttt{Hum}              & \three{\underline{84.4}}{\underline{75.3}}{182.9} & \three{66.7}{74.6}{176.7} & \three{45.5}{73.1}{194.3} & \three{30.4}{71.6}{\underline{137.1}} & \three{\textbf{95.2}}{\textbf{83.2}}{\textbf{129.4}} \\
\texttt{Laughter}         & \three{\underline{47.4}}{68.8}{168.2} & \three{44.9}{74.5}{\underline{120.0}} & \three{42.1}{\underline{75.0}}{153.7} & \three{20.1}{71.8}{128.0} & \three{\textbf{83.7}}{\textbf{78.3}}{\textbf{112.6}} \\
\texttt{Lip Smack}        & \three{\underline{10.8}}{\textbf{66.9}}{\textbf{32.0}}   & \three{6.8}{\underline{59.1}}{\underline{47.2}}   & \three{3.2}{58.5}{166.0}  & \three{0.0}{--}{--}     & \three{\textbf{21.1}}{58.7}{73.7} \\
\texttt{Moan}             & \three{\underline{71.4}}{\underline{77.6}}{70.1}  & \three{26.7}{76.4}{79.4}  & \three{4.8}{70.1}{\underline{45.5}}   & \three{0.0}{--}{--}     & \three{\textbf{74.6}}{\textbf{86.0}}{\textbf{41.3}} \\
\texttt{Question-X}       & \three{\underline{42.0}}{70.4}{54.8}  & \three{10.5}{73.8}{73.0}  & \three{25.0}{\underline{77.5}}{\underline{48.5}}  & \three{0.0}{--}{--}     & \three{\textbf{61.9}}{\textbf{78.0}}{\textbf{39.5}} \\
\texttt{Roar}             & \three{\textbf{88.5}}{77.8}{95.2}  & \three{\underline{67.9}}{\underline{80.1}}{85.0}  & \three{37.2}{74.4}{141.3} & \three{12.1}{63.0}{\underline{83.0}}  & \three{35.9}{\textbf{83.9}}{\textbf{62.1}} \\
\texttt{Scream}           & \three{\textbf{82.2}}{75.3}{\underline{93.9}} & \three{\underline{70.4}}{\underline{75.5}}{99.2}  & \three{41.7}{71.2}{134.4} & \three{39.0}{68.0}{133.8} & \three{40.0}{\textbf{91.0}}{\textbf{27.0}} \\
\texttt{Shush}            & \three{\underline{72.2}}{77.8}{66.9}  & \three{63.3}{76.3}{62.1}  & \three{56.3}{75.9}{\underline{59.8}}  & \three{24.4}{\underline{79.0}}{62.4}  & \three{\textbf{91.2}}{\textbf{87.2}}{\textbf{28.7}} \\
\texttt{Sigh}             & \three{\underline{38.8}}{79.2}{78.9}  & \three{24.1}{\underline{81.7}}{\underline{65.8}}  & \three{33.8}{77.9}{77.1}  & \three{13.1}{69.3}{101.0} & \three{\textbf{84.6}}{\textbf{85.1}}{\textbf{58.9}} \\
\texttt{Sneeze}           & \three{\underline{62.3}}{74.6}{\underline{87.9}}  & \three{59.6}{\underline{76.6}}{108.8} & \three{29.5}{67.0}{113.8} & \three{14.5}{67.7}{110.3} & \three{\textbf{89.3}}{\textbf{85.4}}{\textbf{47.0}} \\
\texttt{Sniff}            & \three{\underline{33.9}}{\underline{72.8}}{66.6}  & \three{33.3}{69.3}{76.3}  & \three{22.1}{67.3}{78.1}  & \three{8.9}{65.1}{\underline{51.7}}   & \three{\textbf{85.0}}{\textbf{82.9}}{\textbf{40.7}} \\
\texttt{Snore}            & \three{\underline{68.6}}{\underline{71.4}}{111.3} & \three{36.7}{68.2}{190.1} & \three{37.5}{66.4}{197.2} & \three{20.0}{64.1}{\underline{78.0}}  & \three{\textbf{95.4}}{\textbf{84.5}}{\textbf{46.7}} \\
\texttt{Surprise-X}       & \three{24.7}{69.2}{76.1}  & \three{18.2}{\textbf{84.8}}{\textbf{30.2}}  & \three{\underline{40.8}}{67.9}{85.2}  & \three{3.0}{70.8}{\underline{32.0}}   & \three{\textbf{70.8}}{\underline{79.4}}{48.3} \\
\texttt{Throat Clearing}  & \three{\underline{32.1}}{70.8}{\underline{69.7}}  & \three{21.5}{\underline{73.5}}{\textbf{56.3}}  & \three{32.0}{68.7}{85.1}  & \three{8.9}{72.6}{73.5}   & \three{\textbf{82.2}}{\textbf{75.9}}{74.9} \\
\texttt{Whistle}          & \three{\underline{65.6}}{68.5}{104.0} & \three{30.4}{\textbf{79.7}}{\underline{72.9}}  & \three{29.2}{67.3}{129.9} & \three{11.1}{69.1}{77.8}  & \three{\textbf{96.9}}{\underline{77.5}}{\textbf{44.2}} \\
\texttt{Yawn}             & \three{\underline{71.2}}{75.5}{137.6} & \three{25.8}{\underline{86.4}}{\underline{83.6}}  & \three{11.8}{67.5}{166.3} & \three{6.7}{52.8}{154.5}  & \three{\textbf{82.6}}{\textbf{88.0}}{\textbf{60.8}} \\
\midrule
Micro avg.                  & \three{\underline{45.8}}{73.6}{\underline{99.0}}  & \three{30.1}{\underline{74.6}}{104.9} & \three{28.8}{70.9}{104.4} & \three{16.7}{69.2}{100.3} & \three{\textbf{71.0}}{\textbf{81.9}}{\textbf{60.6}} \\
Macro avg.                  & \three{\underline{49.8}}{72.8}{\underline{92.6}}  & \three{32.9}{\underline{73.3}}{100.0} & \three{27.9}{69.9}{112.6} & \three{12.5}{69.7}{98.3}  & \three{\textbf{70.2}}{\textbf{80.4}}{\textbf{59.6}} \\
\bottomrule
\end{tabular}%
}
\vspace{5pt}
\end{table}

\subsection{Fine-Grained Category Analysis}
Despite the strong aggregate results, Table~\ref{tab:main_results} reveals considerable variation across the 26 NVV categories, reflecting differences in acoustic saliency and boundary ambiguity. \texttt{Whistle} (96.9\% F1, 44.2\,ms mMAE), \texttt{Cough} (78.1\%, 50.0\,ms), \texttt{Sneeze} (89.3\%, 47.0\,ms), and \texttt{Shush} (91.2\%, 28.7\,ms) achieve high Event-Level F1 with relatively precise boundaries. Their abrupt energy transitions and distinctive harmonic or broadband spectral patterns provide clear cues for both event identification and onset--offset estimation.

In contrast, sustained or repetitive events can achieve high Event-Level F1 while exhibiting larger boundary errors. \texttt{Laughter} (83.7\% F1) and \texttt{Hum} (95.2\% F1) yield mMAEs of 112.6\,ms and 129.4\,ms, respectively. Their periodic and gradually evolving acoustic structure makes the exact onset and offset less sharply defined. \texttt{Applause} (90.6\% F1, 81.1\,ms) and \texttt{Yawn} (82.6\%, 60.8\,ms) show a similar but less pronounced pattern.

Brief, low-energy, and speech-adjacent events remain more challenging. \texttt{Lip Smack} (21.1\% F1) is typically an ultra-short transient, often shorter than 50\,ms, and can be easily masked by reverberation or background noise. \texttt{Gasp} (16.7\%) and \texttt{Breath} (47.7\%) often exhibit diffuse broadband energy with weak harmonic structure, making them less distinguishable from the local noise floor. Moreover, respiratory events frequently occur near lexical boundaries and may blend with neighboring speech or articulatory preparation, further increasing onset--offset ambiguity. In contrast, \texttt{Sniff} achieves 85.0\% F1 with a 40.7\,ms mMAE, suggesting that its characteristic spectral profile provides more stable discriminative cues. Overall, the category-wise results indicate that performance is shaped by acoustic saliency, event duration, and boundary ambiguity.

\subsection{Ablation Study: Validating the Data Construction Pipeline}

To quantify the contribution of each pipeline component, we remove Dual-LLM Verification (Phase 1), Energy-Based Boundary Refinement (Phase 3), and Decoupled Data Augmentation (Phase 4), while keeping the model architecture and training protocol unchanged. Table~\ref{tab:ablation} reports the results. We emphasize Macro F1 because it weights all categories equally and is more sensitive to degradation in sparse NVV types; mIoU and mMAE further assess temporal boundary accuracy.

\begin{table}[htbp]
\centering
\caption{Ablation results on NVV-TimeBench. Each variant removes one component of the automatic data construction pipeline while keeping the downstream model architecture and training protocol unchanged.}
\vspace{-8pt}
\label{tab:ablation}
\begin{tabular*}{\textwidth}{@{\extracolsep{\fill}} l c c c @{}}
\toprule
\textbf{Model Configuration} & \textbf{Macro F1(\%)} & \textbf{mIoU(\%)} & \textbf{mMAE(ms)} \\
\midrule
\textbf{NVV-Locator (Full Pipeline)}  & \textbf{70.2} & \textbf{80.4} & \textbf{59.6} \\
\midrule
w/o Dual-LLM Filtering (Phase 1)     & 55.1 & 74.9 & 73.2 \\
w/o Energy Refinement (Phase 3)      & 51.7 & 62.5 & 133.6 \\
w/o Decoupled Augmentation (Phase 4) & 66.4 & 78.3 & 67.5 \\
\bottomrule
\end{tabular*}
\end{table}

\subsubsection*{Impact of Dual-LLM Verification (Phase 1)}
Removing Dual-LLM Verification reduces Macro F1 from 70.2\% to 55.1\%, decreases mIoU from 80.4\% to 74.9\%, and increases mMAE from 59.6\,ms to 73.2\,ms. Without this quality-control stage, samples with incorrect labels, background interference, overlapping speech, or acoustically ambiguous vocalizations are more likely to remain in the training set, especially for sparse categories. Such noisy supervision can introduce spurious acoustic--label associations and weaken both category discrimination and boundary learning. The 15.1-point Macro F1 drop highlights the importance of reliable category supervision.

\vspace{-10pt}
\subsubsection*{Impact of Energy-Based Boundary Refinement (Phase 3)}
Removing Energy-Based Boundary Refinement causes the largest localization degradation: mMAE rises from 59.6\,ms to 133.6\,ms, mIoU falls from 80.4\% to 62.5\%, and Macro F1 drops to 51.7\%. The transcript-guided windows from LLM-ForcedAligner identify the lexical gap around an NVV but may include silence, background noise, or speech-adjacent content, thereby overestimating the true event extent and providing imprecise onset--offset supervision. Under the Event-Level F1 criterion, category-correct predictions with t-IoU below $0.5$ are counted as unmatched, affecting both precision and recall. Energy-based refinement is therefore essential for converting coarse lexical gaps into acoustically grounded NVV boundaries.

\vspace{-9pt}
\subsubsection*{Impact of Decoupled Data Augmentation (Phase 4)}
Removing Decoupled Data Augmentation yields a smaller but consistent decline: Macro F1 decreases to 66.4\%, mIoU to 78.3\%, and mMAE increases to 67.5\,ms. Phase 4 supplements each refined sample with the original mixed speech--NVV recording, an isolated NVV segment, and an NVV-removed speech context, while increasing diversity for underrepresented categories through class balancing. These complementary views help distinguish genuine NVVs from lexical speech and speech-adjacent acoustic variation, providing useful additional supervision beyond the original mixed recordings alone.

\subsection{Cross-Corpus Generalization Evaluation}
We further evaluate NVV-Locator on WESR-Bench to assess its cross-corpus transferability. WESR-Bench contains over 900 utterances with diverse vocal events. After mapping its labels to our unified taxonomy, NVV-Locator is evaluated zero-shot without dataset-specific fine-tuning, achieving a Micro F1 of 70.2\% and a Macro F1 of 45.7\%.
For reference, the WESR-trained systems reported in the original WESR-Bench study achieve Micro F1 scores of 70.5--71.4\% and Macro F1 scores of 37.7--38.0\%. Although a strict comparison is not possible because our evaluation involves taxonomy mapping and no target-corpus training, the results are of a comparable overall scale. In particular, the zero-shot performance suggests that NVV-Locator retains effective NVV event recognition ability beyond the NVV-TimeBench distribution.

\section{Conclusion}
In this paper, we present a scalable approach to fine-grained nonverbal vocalization (NVV) temporal grounding, covering taxonomy construction, automatic timestamp annotation, benchmark curation, and model formulation. We unify 26 NVV categories and develop a four-stage data construction pipeline that combines dual-LLM verification, transcript-guided alignment, energy-based boundary refinement, and decoupled augmentation. We further introduce NVV-TimeBench and propose NVV-Locator, a non-autoregressive slot-filling model for joint NVV recognition and waveform-time localization. Experimental results show that the proposed approach substantially outperforms the evaluated large audio model counterparts, while ablation studies validate the importance of the data construction pipeline for both event recognition and boundary localization. Evaluation on an external corpus further supports its cross-corpus generalization.

\bibliographystyle{splncs04}
\bibliography{cya_google}

\end{document}